\documentstyle[twocolumn,aps]{revtex}

\begin{document}

\hyphenation{Kri-sha-na}

\twocolumn[
\hsize\textwidth\columnwidth\hsize\csname@twocolumnfalse\endcsname
\draft

\title{Dissipation effects on the
superconductor-insulator transition in 2-D superconductors}
\author{N. Mason and A. Kapitulnik}
\address{Departments of Applied Physics and of Physics, Stanford University,
Stanford, CA 94305, USA}
\date{\today}
\maketitle

\begin{abstract}
	Results on the superconductor to
insulator transition in two-dimensional films are analyzed in terms of
coupling of the system to a dissipative bath. Upon lowering
the temperature the parameter that controls this coupling becomes
relevant and a wide range of metallic phase is recovered.
\end{abstract}

\pacs{PACS numbers: 74.20.-z, 74.76.-W, 73.40.Hm }
]

	Quantum phase transitions (QPT) continue to attract intense
theoretical and
experimental interest. These transitions -- where changing an external
parameter in the Hamiltonian of the system induces a transition from one
quantum ground state to another, fundamentally different one -- have
been invoked to explain data from various experiments, including
quantum-Hall liquid to insulator, metal to insulator and superconductor to
insulator experiments. In this paper we use experiments done on thin
superconducting films which undergo a so-called Superconductor-Insulator (SI)
transition to propose a new phase diagram for a QPT that includes a
dissipation axis. The implications of our analysis go much beyond this problem
and in fact bear on all QPT in two dimensions.

	The most common treatment of the SI transition is to map it onto the
so-called  ``dirty-bosons'' model. In this model Cooper pairs are the
Bose particles with well formed (non-fluctuating) pair  amplitudes. At
zero temperature an SI transition is expected as a function of disorder,
with vortex-antivortex pairs activated by both quantum fluctuations and
disorder. Vortices of one vorticity are also induced by a magnetic field;
with no disorder an
Abrikosov lattice of these vortices realizes the superconducting phase.
However, an arbitrary amount of disorder disrupts the lattice and a true
superconducting phase is assumed to be recovered only at T=0. In the
superconducting phase vortices are localized into a so-called vortex-glass
phase \cite{fisher1} and the Cooper pairs are delocalized. Upon increasing
the magnetic field above some critical field $H_c$ , vortices will delocalize
with Cooper pairs localizing into a Bose-Glass phase. Further increase of the
magnetic field dissociates the Cooper pairs, and fermion  degrees of freedom
then determine the properties of the system. At the  SI transition, both
vortices and Cooper pairs are delocalized in a ``Bose-Metal'' phase;
thus, a finite conductivity is expected at the transition. The above
scenario was the basis for a scaling theory proposed by Fisher
\cite{fisher1} in which a field-tuned transition was considered as a
continuous transition with an associated diverging length
\( \xi\!\sim\!(H_c-H)^{-\nu}\), as well as a finite, universal conductivity
at the transition \cite{fisher2}.

	The above ``dirty-bosons'' model was used to describe numerous
experimental data in both disorder and field tuned transitions
 \cite{yazdani,haviland,hebard}. However, a difficulty in applying this
model arose when the vortex glass phase displayed a temperature
independent resistance that could be interpreted as quantum tunneling of
vortices \cite{ephron,liu,zant}. In particular, in previous experiments on
 amorphous MoGe \cite{yazdani} we showed scaling
for a range of disorder; yet the superconducting phase at lower
fields  showed finite resistance
indicative of quantum tunneling of  vortices
\cite{ephron}. Assuming continuity as a function of
increasing magnetic field, we  concluded that there was
no true superconducting phase all the  way
to the critical field. A closer inspection of the data presented  in
\cite{yazdani} and
\cite{ephron} suggests that while the quantum critical point is apparent at low
temperatures with  excellent agreement with scaling, lowering the
temperature  further results in a rapid disruption of scaling. We
interpret these results as evidence of a coupling of our system to a
dissipative environment, presumably a background of delocalized fermions
as was first conjectured by Yazdani and Kapitulnik \cite{yazdani}.

 Such coupling to dissipation could lead to a new phase diagram
for the system, shown pictorially in figure 1. To the standard H-T-disorder
phase diagram \cite{fisher1}, we add a new axis, 
$\alpha$, representing the strength of the
coupling to a dissipative bath (so the figure is a slice at fixed
disorder).  Upon cooling the system the coupling to the bath
 becomes more relevant and the  system flows away from the
unstable critical point (the pure dirty bosons point) into a wide range of
``metallic'' behavior. In fact, depending on the system, the
opening  of the metallic region could leave a diminishingly small
region where  true superconductivity can be found. In the case of MoGe
with $R_\Box$  of order 1.5 k$\Omega$ we showed \cite{ephron} that
already at 20\% of $H_c$, superconductivity is lost.

	Continuing to use amorphous-MoGe films as our model system,
we have conducted further measurements to better examine the low
temperature ``metallic'' behavior and dissipative coupling. These
experiments were performed on thin films grown by multitarget magnetron
sputtering on a SiN substrate with a Ge buffer layer. The films were
grown in the same sputtering runs as those used in \cite{ephron} and
\cite{yazdani}; details of growth and characterization are described
elsewhere \cite{yoshizumi}. Most of the data reported in this paper was taken
on films with $x=$0.43, thickness of 30 $\AA$ and Tc $\sim$0.5 K. Previous
studies have determined the films to be highly amorphous and homogeneous
over all relevant length scales.
 The films were patterned into 4-probe structures, and measured in a
dilution refrigerator using standard  low-frequency lock-in techniques. Care
was taken to eliminate spurious noise and heating effects. A typical set of
resistance vs. temperature for increasing magnetic field data is given
in figure 2. Similar results were obtained for all films including those
reported in \cite{yazdani,ephron}. A main feature of these result is that upon
lowering the temperature the activated behavior of the resistance
changes to a temperature independent resistance as the temperature
approaches zero, as can be seen in the log $R_\Box$ vs. 1/T inset of figure
2. In \cite{ephron} the low temperature  saturation value of the resistance
obeys the empirical form
\begin{equation}
R(H) = {R_0} e{^{c( {\hbar} /{e^2}R_\Box)H/H_{c2}}},
\end{equation}
with $R_\Box$ being the resistance per square of the sample and
c$\simeq$2 is a constant. This unusual behavior was explained by
Shimshoni {\it et al.} \cite{shimshoni} using dissipative quantum
tunneling  of vortices from one ``insulating'' puddle to its neighbor.
The source of dissipation was assumed to be the electrons in the core of the
vortex, suggesting the use of Bardeen-Stephen \cite{bardeen} expression for the
viscosity in the Euclidian action governing the tunneling. If the vortex
tunneling is mediated by coupling to a dissipative bath, then finite diffusion
appears which explains the flattening of the resistance as T approaches zero.
This model produces  excellent fit to the experimental data (e.g., the
theoretical model, using  Bardeen-Stephen dissipation gives
c$\simeq$1.6 for the prefactor). Since the superconducting phase
is obtained in this model by percolation of couplings of superconducting
puddles, this model also explains the exponent $\nu \simeq$1.35 found in all
field-tuned transitions by many groups since the correlation-length
exponent in 2-D classical  percolation is 4/3.

	To check the scaling and flattening of the resistance in the MoGe
films, we extended the temperature range by  measuring down to 20 mK
from low field through the presumed SI transition. A blow-up of the
transition region is shown in figure 3. Here we observe that as we go
to low temperatures, all curves, whether initially decreasing or
increasing, flatten. This affects scaling in a dramatic way. 
Following  \cite{fisher1} we fit isotherms for T
$\ge$  100 mK to:$R = R_c{\cal{F}}[(H-H_c)/{T^{1/z\nu}}]$. 
The scaling function $\cal{F}$$(x)$
displays two branches,  for positive (``insulating'') and negative
(``superconducting'') arguments. A best fit to the
scaling function for high temperature isotherms gives $z\nu = 1.33\pm 0.05$,
as shown in the upper part of figure 4. The dotted line in the figure shows
the deviation of the 50mK line from the other scaled curves. This
deviation, which evinces the breakdown of scaling with respect to the
critical point, is amplified in the lower part of figure 4.  We believe
that this low temperature deviation from scaling and flattening of the
resistance are a manifestation that the isolated metallic point at $H_c$
``opens-up'' to a region of metallic behavior as a function of some parameter
that becomes relevant at low temperatures. Adding the hypothesis of vortex
tunneling in a dissipative medium we conclude that
the natural parameter that ``pulls'' the system
away from the SI transition critical point is the strength of the coupling
to the dissipative bath. Starting with a system with fixed disorder we
naturally arrive at the phase diagram presented in figure 1. Note that the
phase diagram allows for a finite range of $\alpha$ in which the SI
transition is preserved, a feature that we discuss next.

	Dissipation in connection with the S-I
transition was first studied by Chakravarty {\it et al.}
 \cite{chakravarty1,chakravarty2} with a quantum statistical
mechanics model of an array of
resistively shunted Josephson junctions. This model predicted a
superconducting to normal-state transition as a function of dissipation.
More recently Wagenblast {\it et al.} \cite{wagenblast}
attempted to explain non-universal results observed in experiments
on homogeneous \cite{yazdani} and inhomogeneous thin films
\cite{haviland} by introducing a model with local dissipation in which the
dissipative term couples only to the phase of a single island. Their model
led to a new  universality class at the superconductor-insulator
transition. In particular, they found that the conductivity at criticality
is non-universal and is characterized by a damping-dependent dynamical
critical exponent. Both models \cite{chakravarty2,wagenblast} preserve the
SI transition but do not allow for a range of a metallic phase.
Along with the fact that neither model includes disorder (in
\cite{chakravarty1}
randomness was discussed and argued not to change the nature of the phase
diagram), we believe that the absence of dual
excitations (i.e. vortices) in the treatment of the
 transition excludes a finite range of metallic phase.
In the absence of vortices, a metallic phase is excluded
automatically, according to our proposed scenario as outlined below
\cite{debo}.
However, in the absence of dissipation and for strong
enough disorder a pure SI transition is
expected according to \cite{fisher1} and thus we allow for a range,
$\alpha$ $<$ $\alpha_c$, for which a metallic phase
exists only at the transition point itself.
 For weak disorder this range may be very small, as is presumably
the case for the MoGe films discussed here.

	Levelling of the resistance -- and therefore a probable coupling to
dissipation -- in field tuned transitions has been observed in other
experiments on thin  superconducting films \cite{liu} and in Josephson
Junctions arrays \cite{zant}. Even more intriguing is the fact  that the
analogous problem of the quantum Hall effect to insulator transition
(QHIT) exhibits similar effects. In this  problem a sharp change in the
behavior of the resistivity at low temperature as a function of the magnetic
field has been interpreted  as a quantum phase transition between localized
bosons and localized vortices \cite{fisher3}. While scaling has been observed
with high accuracy for this problem \cite{sondhi}, recent experiments reveal
that occasionally samples fail to exhibit scaling, and the resistance on
the quantum Hall liquid side levels to a constant, suggesting quantum
tunneling \cite{shahar}.

	Quantum tunneling of vortices in the SI transition problem, or of edge
states in the QHIT problem, are a natural consequence of the Shimshoni {\it
et al.} model \cite{shimshoni}. This model predicts continuous transitions for
the SI transition or QHIT cases which are percolation-like. Its importance is
that it allows for a destruction of the superfluid phase  due to incoherent
tunneling of vortices (in the SI transition) or of edge states (in the QHIT).
The missing ingredient of the model is the existence of a dissipative bath to
which the system may couple with some strength $\alpha$. This deficiency,
which is corrected in the present paper, leads to the
 generalized phase diagram presented in figure 1. Combining the
percolation model of \cite{shimshoni}, the argument of
\cite{chakravarty1} that in the random array model the transition
is governed by the properties of the typical junction, and the argument of
Schmid
\cite{schmid} that depending on the parameters of the junction
its behavior will change from coherent to diffusive, it seems that a
metallic phase
must exist in the SI problem. A similar argument should hold for the QHIT
problem.
However, for samples in which the dissipation is weak the QHIT
may still be preserved, as has been observed by many authors
\cite{listqhe}. In these cases it is no surprise that the
resistance at the critical field is
the quantum resistance, $h/e^{2}$, and that the exponent is
$\nu$$\sim$7/3, a value consistent with quantum percolation.
 The failure to observe scaling in samples that show flattening
of the resistance \cite{shahar} is interpreted as an indication
that a stronger $\alpha$ exist in these samples.
We therefore suggest that for these samples scaling
should be attempted with the high
temperatures data only. This should lead to a result similar to the one
presented here for the SI transition with $\nu$$\sim$4/3. In fact, following
our prediction we examined figure 3(a) of \cite{shahar}, which presented
 the asympthotical behavior of the activation energy in the QH state.
This activation energy is proportional to
$T^{1/z\nu}$ and gives $1/z\nu $$\simeq$ 3/4, very different
from other QHIT results of 3/7 or 3/14. While scaling is observed in this
high temperature limit, the critical resistance is different from the
universal resistance, similar, again, to the SI results. Dephasing in the
QHIT problem on the insulating side was discussed by Pryadko and Auerbach
\cite{pryadko} resulting in a finite Hall
resistivity at T=0. However, the source of dissipation
is still unknown in this problem.

	Finally we comment that our discussion above is very general and should
apply to any two-dimensional quantum critical point where dissipation
can be relevant. A possible candidate is the newly discovered metal-insulator
transition in Si MOSFETs. Here the insulator is believed to be a Wigner
crystal. In the presence of disorder this problem resembles that of a vortex
glass in the SI transition problem.

	In conclusion, in this paper we presented a new phase diagram for the
superconductor-insulator transition problem in the presence of coupling to a
dissipative bath. We argued that as the critical point is approached by e.g.
lowering the temperature, this coupling may become a relevant variable in the
problem and thus pull the system to a new phase which is metallic in nature.
We further proposed that this scenario is very general and may explain the
recent results on flattening of the resistance observed in
quantum Hall to insulator transitions.

	We thank Assa Auerbach, Steve Kivelson, Sudip Chakravarty, Seb
Doniach and Debopriya Das for many useful discussions. Work supported by NSF grant
DMR-9800663. NM thanks NSF for fellowship support. Samples prepared at
Stanford's Center for Materials Research.

%

\figure{FIG. 1. Phase diagram for the field-tuned SI transition:
 H-T-Dissipation strength diagram at finite disorder.
$\alpha_c$ marks the point at which a finite range of metallic phase opens
up.
\label{fig1}}

\figure{FIG. 2.
Set of resistance vs. temperature curves for
B=2,1.3,1.23,1.18,1.1,1.0,.75,.70,0 Tesla Inset: log $R_\Box$ vs 1/T for
0,.6,.7,.75,1,1.18,1.21 Tesla
\label{fig2}}

\figure{FIG. 3.
Set of resistance vs. temperature curves for B=1.3,1.26,1.23,1.21,1.2,1.18
Tesla. Flattening of the resistance at low temperatures coincides with
breakdown of scaling.
\label{fig3}}

\figure{FIG. 4.
Upper: Scaling for T $\ge$ 100 mK yielding: z$\nu$=1.33$\pm$0.05.
50mK curve (dotted line) strongly deviates from others.
Lower: Deviation of scaled curves from ``standard'' (100mk) curve. Note that
only the low temperature curves show significant deviation.
\label{fig4}}

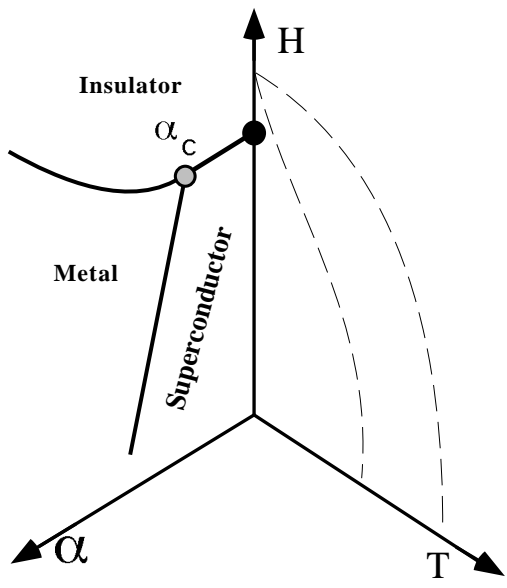
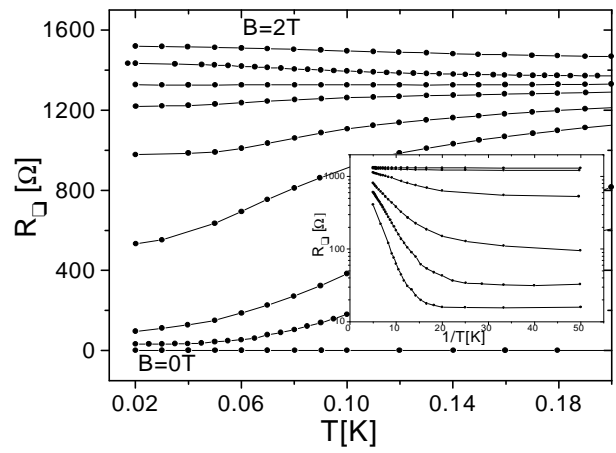
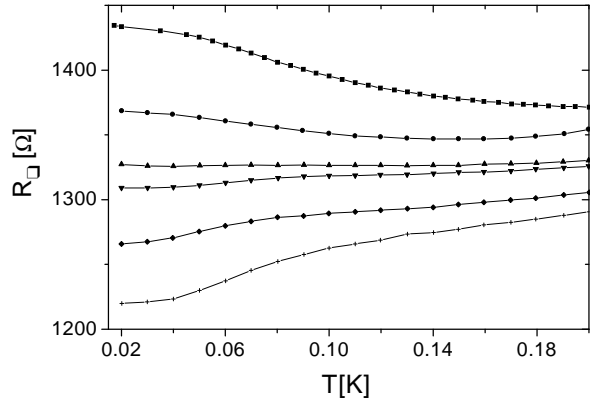
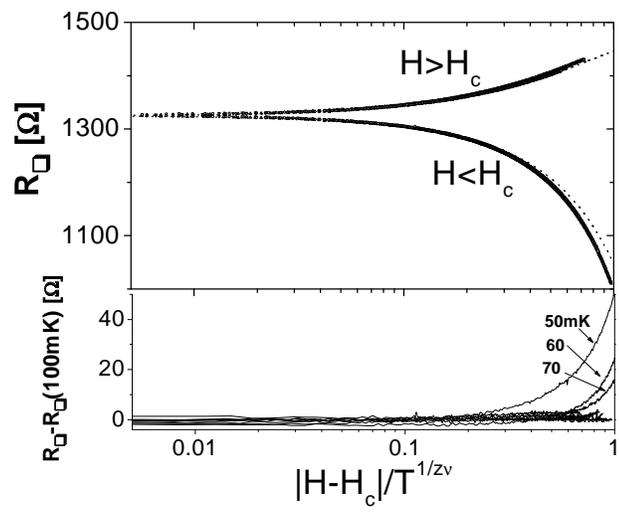
\end{document}